\documentclass{emulateapj}

\slugcomment{Accepted for publication in the Special GALEX ApJ Supplement, December 2007}
\shortauthors{Ree et~al.}
\shorttitle{Look-back Time Evolution of Far-UV Flux from Brightest Cluster Ellipticals}

\begin{document}

\title{The Look-back Time Evolution of Far-Ultraviolet Flux from the Brightest Cluster Elliptical Galaxies at $\lowercase{z}<0.2$}

\author{
Chang~H.~Ree\altaffilmark{1},
Young-Wook~Lee\altaffilmark{1,2},
Sukyoung~K.~Yi\altaffilmark{1},
Suk-Jin~Yoon\altaffilmark{1},
R.~Michael~Rich\altaffilmark{3},
Jean-Michel~Deharveng\altaffilmark{4},
Young-Jong~Sohn\altaffilmark{1},
Sugata~Kaviraj\altaffilmark{5},
Jonghwan~Rhee\altaffilmark{1,3},
Yun-Kyeong~Sheen\altaffilmark{1},
Kevin~Schawinski\altaffilmark{5},
Soo-Chang~Rey\altaffilmark{6},
Alessandro~Boselli\altaffilmark{4},
Jaehyon~Rhee\altaffilmark{1,7},
Jose~Donas\altaffilmark{4},
Mark~Seibert\altaffilmark{7},
Ted~K.~Wyder\altaffilmark{7},
Tom~A.~Barlow\altaffilmark{7},
Luciana~Bianchi\altaffilmark{8},
Karl~Forster\altaffilmark{7},
Peter~G.~Friedman\altaffilmark{7},
Timothy~M.~Heckman\altaffilmark{9},
Barry~F.~Madore\altaffilmark{10},
D.~Christopher~Martin\altaffilmark{7},
Bruno~Milliard\altaffilmark{4},
Patrick~Morrissey\altaffilmark{7},
Susan~G.~Neff\altaffilmark{11},
David~Schiminovich\altaffilmark{12},
Todd~Small\altaffilmark{7},
Alex~S.~Szalay\altaffilmark{9}, and
Barry~Y.~Welsh\altaffilmark{13}
}

\altaffiltext{1}{Center for Space Astrophysics and Department of Astronomy, Yonsei University, Seoul 120-749, Korea (e-mail : chr@csa.yonsei.ac.kr, ywlee@csa.yonsei.ac.kr)}
\altaffiltext{2}{Department of Astronomy, Yale University, New Haven, CT 06511}
\altaffiltext{3}{Department of Physics and Astronomy, University of California at Los Angeles, Los Angeles, CA 90095}
\altaffiltext{4}{Laboratoire d'Astrophysique de Marseille, BP 8, Traverse du Siphon, 13376 Marseille Cedex 12, France}
\altaffiltext{5}{Department of Physics, University of Oxford, Oxford OX1 3RH, UK}
\altaffiltext{6}{Department of Astronomy and Space Science, Chungnam National University, Deajeon, 305-764, Korea}
\altaffiltext{7}{California Institute of Technology, MC 405-47, 1200 East California Boulevard, Pasadena, CA 91125}
\altaffiltext{8}{Center for Astrophysical Sciences, Johns Hopkins University, 3400 North Charles St., Baltimore, MD 21218}
\altaffiltext{9}{Department of Physics and Astronomy, The Johns Hopkins University, Homewood Campus, Baltimore, MD 21218}
\altaffiltext{10}{Observatories of the Carnegie Institution of Washington, 813 Santa Barbara St., Pasadena, CA 91101}
\altaffiltext{11}{Laboratory for Astronomy and Solar Physics, NASA Goddard Space Flight Center, Greenbelt, MD 20771}
\altaffiltext{12}{Department of Astronomy, Columbia University, New York, NY 10027}
\altaffiltext{13}{Space Sciences Laboratory, University of California at Berkeley, 601 Campbell Hall, Berkeley, CA 94720}

\begin{abstract}
We present the GALEX UV photometry of the elliptical galaxies in Abell clusters at moderate redshifts $(z<0.2)$ for the study of the look-back time evolution of the UV upturn phenomenon. The brightest elliptical galaxies (M$_r\lesssim-22$) in 12 remote clusters are compared with the nearby giant elliptical galaxies of comparable optical luminosity in the Fornax and Virgo clusters. The sample galaxies presented here appear to be quiescent without signs of massive star formation or strong nuclear activity, and show smooth, extended profiles in their UV images indicating that the far-UV (FUV) light is mostly produced by hot stars in the underlying old stellar population. Compared to their counterparts in nearby clusters, the FUV flux of cluster giant elliptical galaxies at moderate redshifts fades rapidly with $\sim2$~Gyrs of look-back time, and the observed pace in $FUV-V$ color evolution agrees reasonably well with the prediction from the population synthesis models where the dominant FUV source is hot horizontal-branch stars and their progeny. A similar amount of color spread ($\sim1$~mag) in $FUV-V$ exists among the brightest cluster elliptical galaxies at $z\sim0.1$, as observed among the nearby giant elliptical galaxies of comparable optical luminosity.
\end{abstract}


\section{INTRODUCTION}

The ultraviolet (UV) upturn phenomenon of early-type galaxies is the rising flux with decreasing wavelength from 2500{\AA} to the Lyman limit. Since its first detection (Code 1969), many UV space facilities have targeted nearby elliptical galaxies and spiral bulges in order to investigate the spectral and photometric characteristics of the UV upturn in early-type galaxies and its connection to the physical properties (Faber 1983; Burstein et~al. 1988; Ferguson et~al. 1991; O'Connell et~al. 1992; Brown et~al. 1997; Ohl et~al. 1998; see O'Connell 1999 and references therein). Although it is now well established that the far-UV (FUV) flux of nearby early-type galaxies originates from a minority population of old hot helium-burning horizontal-branch (HB) stars (e.g., O'Connell 1999; Brown et~al. 2000b), the remaining issue on their metallicities has an outstanding implication on the galaxy evolution (see Yi et~al. 1999 and references therein).

Burstein et~al. (1988) found that the UV upturn strength correlates with the nuclear spectral line index Mg$_{2}$ (and, weakly, to the central velocity dispersion and luminosity). However, recent studies show that the metallicity may not be the sole parameter controlling the UV flux (Ohl et~al. 1998; O'Connell 1999; Deharveng, Boselli, \& Donas 2002; Rich et~al. 2005). Two different HB hypotheses were proposed -- namely the ``metal-poor'' (Lee 1994; Park \& Lee 1997) and ``metal-rich'' (Bressan, Chiosi, \& Fagotto 1994; Dorman, O'Connell, \& Rood 1995; Yi, Demarque, \& Oemler 1998) HB models according to the mean metallicity of dominant FUV source populations. Both models reproduce the observed UV spectra in nearby elliptical galaxies reasonably well, despite requiring significantly different ages for the nearby giant elliptical galaxies (Yi et~al. 1999). An absorption feature in the UV spectra indicating a low surface metallicity of FUV sources has been found in nearby early-type galaxies (Ferguson et~al. 1991; Brown et~al. 1997); yet it may not reflect the stellar interior abundance due to the heavy element redistribution in the atmospheres of hot evolved stars (Behr et~al. 1999; Moehler et~al. 2000). 

The average temperature of the helium burning stars is mostly controlled by the envelope mass, M$_{env}$, at helium ignition (i.e., being hotter with lower M$_{env}$), while the helium core mass is relatively insensitive to other physical parameters (Iben \& Rood 1970; Rood 1973; Sweigart 1987). The observed strength of FUV flux from local elliptical galaxies implies that the dominant FUV sources are the hot helium burning stars with very small ($< 0.05$ M$_{\odot}$) envelope mass (see O'Connell 1999). At such high temperature region (T$_{eff} \gtrsim 15000~K$), a small change in the envelope mass could induce a rapid evolution in the HB mean temperature (Dorman, Rood, \& O'Connell 1993; Bressan et~al. 1994; Yi, Demarque, \& Oemler 1997; Yi et~al. 1999). Therefore, the temperature of hot HB stars declines rapidly as their envelope mass (i.e., RGB progenitor mass) increases with decreasing age (e.g., Lee, Demarque, \& Zinn 1994), and hence, the UV flux is expected to fade away with look-back time in old stellar systems (Greggio \& Renzini 1990, 1999; Bressan et~al. 1994; Tantalo et~al. 1996; Lee et~al. 1999). Observations for the amplitude and the evolutionary path of the UV flux from elliptical galaxies at moderate redshifts would therefore provide valuable constraints to the model input parameters. Population synthesis models of Yi et~al. (1999) and Yoon (2002) indicate that careful observations for the UV look-back time evolution could also discriminate the two alternative HB solutions on the origin of the UV upturn phenomenon.

Such UV fading at moderate redshifts has been detected by Lee et~al. (2005a), from the \textit{Galaxy Evolution Explorer} (GALEX) observations of the early-type galaxies in the Fornax cluster and Abell~2670 ($z=0.076$). Although the observed fading of the UV upturn is consistent with the variation predicted by the models, it would be premature to conclude that GALEX has detected the look-back time evolution as the result was based only on two clusters at relatively low redshifts. In this paper, as an extension of Lee et~al. (2005a), we present the GALEX UV photometry of the elliptical galaxies in 12 clusters at $0<z<0.2$. We focus on the brightest (M$_r\lesssim-22$) cluster ellipticals (often called ``first-ranked'' ellipticals) that are the most massive galaxies in the Universe and likely to be at the centers of clusters. As inferred from the observations for the nearby galaxy samples (Burstein et~al. 1988; Boselli et~al. 2005; Donas et~al. 2007), they are also most likely the strongest UV emitters among the \textit{quiescent} early-type galaxies in each redshift bin. Recent studies suggest that the most massive (log~$\sigma \sim 2.5$) bright elliptical galaxies in dense cluster environments are relatively free from residual star formation and AGN accretion flow (Bower et~al. 2006; Jimenez et~al. 2006; Schawinski et~al. 2006, 2007; Kauffmann et~al. 2007). Investigations for the less massive early-type galaxies in clusters are ongoing, which requires the optical spectroscopic verification with deeper survey depths than the \textit{Sloan Digital Sky Survey} (SDSS) magnitude limit ($r<17.77$).

Since our first report on the GALEX look-back time evolution of UV upturn (Lee et~al. 2005a), a new paradigm has been suggested on the origin of hot HB stars in old stellar systems. This new theory is based on the recent observations and modeling of some peculiar globular clusters such as $\omega$~Cen and NGC 2808, where the special features on the main-sequence and hot HB in these clusters can only be explained by the presence of super-helium-rich subpopulations (D'Antona \& Caloi 2004; Norris 2004; D'Antona et~al. 2005; Lee et~al. 2005b; Piotto et~al. 2005). Although the origin of this helium enhancement is not fully understood yet, the likely presence of super-helium-rich subpopulations and the resulting hot HB stars in old stellar systems deserve further investigation, as they could be a major source of the FUV flux in quiescent elliptical galaxies. The temperature of these helium-rich HB stars, if present in elliptical galaxies, must also decrease as their envelope mass increases with look-back time (Sweigart 1987; Lee et~al. 1994), and therefore we expect similar fading of FUV flux with redshift. The detailed evolution of FUV flux under this scenario, however, requires specific population synthesis models with super-helium-rich subpopulations included, and the present observations will eventually be used to discriminate the suggested scenarios on the origin of hot HB stars in quiescent elliptical galaxies when these new models are available.

\section{OBSERVATIONS AND DATA REDUCTION}

\subsection{GALEX UV imaging}

In order to investigate the origin of the UV upturn phenomenon in early-type galaxies, GALEX has been performing an imaging survey for the elliptical-rich Abell clusters at moderate redshifts ($z<0.2$) in the FUV ($1344 - 1786$~\AA) and near-UV (NUV; $1771 - 2831$~\AA) bandpasses. The GALEX field of view is circular, with a diameter of 1.2 degree, and each image contains 3840~$\times~$3840 pixels with 1.5~arcsec/pixel scale. See Martin et~al. (2005) and Morrissey et~al. (2005) for details on the GALEX instruments and mission.

Using the GALEX GR2 public release and IR1.1 internal release data sets, we have analyzed 7 Deep Imaging Survey (DIS : $6 \sim 30$ ksecs) and 5 Medium Imaging Survey (MIS : $1.5 \sim 3$ ksecs) fields for the Abell clusters at $z = 0.05 \sim 0.17$. Basic information on the target clusters and UV observations is presented in Table~1.

\begin{deluxetable*}{ccccccl}
\tablecaption{GALEX observations}
\tablewidth{0pt}
\tablehead{
\colhead{Target} &
\colhead{$z\tablenotemark{a}$} &
\colhead{RA$\tablenotemark{a}$} &
\colhead{DEC$\tablenotemark{a}$} &
\colhead{E$(B-V)\tablenotemark{a}$} &
\colhead{Tilename} &
\colhead{Exposure} \\
\colhead{cluster} &
\colhead{} &
\colhead{J2000} &
\colhead{J2000} &
\colhead{} &
\colhead{} &
\colhead{FUV / NUV} \\
\colhead{} &
\colhead{} &
\colhead{(h:m:s)} &
\colhead{(d:m:s)} &
\colhead{} &
\colhead{} &
\colhead{(sec)} 
}
\startdata
Abell~2399 & 0.058 & 21:57:32.5 & $-$07:47:40 & 0.039 & MISDR2\_20914\_0716 & \phn2972 / \phn2972\\
Abell~2670 & 0.076 & 23:54:10.1 & $-$10:24:18 & 0.043 & UVE\_A2670          & \phn6876 / \phn6876\\
Abell~2249 & 0.082 & 17:09:43.0 & $+$34:27:18 & 0.025 & MISDR2\_11627\_0974 & \phn1356 / \phn1356\\
Abell~2448 & 0.082 & 22:31:43.6 & $-$08:26:33 & 0.055 & MISDR2\_29594\_0721 & \phn1669 / \phn4868\\
Abell~3330 & 0.092 & 05:14:47.4 & $-$49:04:19 & 0.028 & UVE\_A3330	    & \phn6377 / 27307\\
Abell~2048 & 0.097 & 15:15:17.8 & $+$04:22:56 & 0.048 & UVE\_A2048          & \phn8312 / \phn8312\\
Abell~0389 & 0.113 & 02:51:31.0 & $-$24:56:05 & 0.016 & UVE\_A0389	    & 18600 / 25696\\
Abell~0733 & 0.116 & 09:01:19.2 & $+$55:37:13 & 0.017 & MISDR1\_03125\_0450 & \phn1693 / \phn1693\\
Abell~1406 & 0.118 & 11:53:15.6 & $+$67:53:19 & 0.013 & MISDR1\_00365\_0492 & \phn1698 / \phn1698\\
Abell~0951 & 0.143 & 10:13:54.8 & $+$34:43:06 & 0.012 & UVE\_A0951          & 16324 / 29401\\
Abell~2235 & 0.151 & 16:54:57.9 & $+$40:01:16 & 0.021 & UVE\_A2235          & 24926 / 32647\\
Abell~1979 & 0.169 & 14:51:00.2 & $+$31:16:41 & 0.017 & UVE\_A1979          & 10109 / 19166
\enddata
\tablenotetext{a}{NASA/IPAC Extragalactic Database (http://nedwww.ipac.caltech.edu)}
\label{tab:tab1}
\end{deluxetable*}

The GALEX data pipeline utilizes the SExtractor image analysis package (Bertin \& Arnouts 1996) for detection and photometry of sources in the imaging data with some modifications in the determinations of sky background and detection threshold. The images were calibrated and processed through the standard GALEX pipeline. In order to minimize the blending effect or UV contamination from star-forming dwarf companion galaxies around the target elliptical galaxy, images were reprocessed with some adjustments in detection parameters. After a series of experiments, DEBLEND\_MINCONT = 5e-05 and DETECT\_MINAREA = 8, along with the other standard pipeline parameters, produced good results in crowded regions. 

We adopt MAG\_AUTO magnitudes measured through elliptical apertures (Kron 1980) from the SExtractor photometry as the total magnitudes of the sources in FUV images. The NUV flux of the target elliptical galaxies is measured within the aperture predetermined in the FUV image via ``double-image mode'' of the SExtractor. In general, the NUV images are deeper than the FUV images with higher signal-to-noise ratio, and hence, better for source detection (Wyder et~al. 2005). However, at the same time, the contamination from warm stars ($\sim10,000$~K) of young or intermediate-age population in the surrounding dwarf companions is more serious in the NUV than in the FUV images of crowded region. Since the main targets here are the most massive elliptical galaxies in dense cluster environments, we choose the elliptical aperture predetermined in FUV photometry to derive $FUV - NUV$ color as a measure of the UV spectral shape. Small amounts of aperture correction in $FUV - NUV$ may be required for the comparison with local samples due to the radial color gradients (Deharveng et~al. 1982; O'Connell et~al. 1992; Ohl et~al. 1998; Donas et~al. 2007). Nevertheless, the UV apertures in our photometry are large enough to include most of the UV light from the hot stars responsible for the FUV flux from elliptical galaxies, and hence we apply no aperture corrections in $FUV - V$ color which is a measure of the UV upturn strength.

The photometric errors are computed to include the background contribution to the Poisson errors, and also the flat field variations (0.05 mag or 4.8\% flux in FUV and 0.03 mag or 2.8\% flux in NUV). All the apparent magnitudes are corrected for foreground extinction using Schlegel, Finkbeiner, \& Davis (1998) reddening maps and the extinction law of Cardelli, Clayton, \& Mathis (1989), assuming $R_{V} = 3.1$. The UV extinction has been calculated with the GALEX filter throughputs convolved to the 42 GISSEL (S. Charlot \& L. Tresse 2006, private communication) galaxy template spectra. We have adopted the median of the coefficients from the template fits: $A_{FUV} = 8.24 \times E(B-V)$, and $A_{NUV} = 8.24 \times E(B-V) - 0.67 \times E(B-V)^{2}$. GALEX uses the AB magnitude system of Oke \& Gunn (1983) :
\begin{displaymath}
m = m_0 - 2.5~\log~(count~s^{-1}),
\end{displaymath}
with magnitude zero points $m_0$ = 18.82 (FUV) and 20.08 (NUV).

\subsection{Optical corollary data}

We have used the SDSS DR4 (York et~al. 2000; Adelman-McCarthy et~al. 2005) and the photometric data in Fasano et~al. (2000) for the identification and optical photometry of the target galaxies in remote clusters. For the clusters of galaxies having SDSS photometry, we have first searched for early-type galaxies within 2~Mpc ($10\sim35$~arcmin at target redshifts with the cosmological parameters ($\Omega_M$, $\Omega_\Lambda$, $H_0$) = (0.3, 0.7, 70)) radius from the cluster central position given in Abell, Corwin, \& Olowin (1989), and then selected the possible members by applying the color--magnitude ($g-r$ vs. $r$) relations combined with the photometric redshifts of Blanton et~al. (2003). From the SDSS $r$-band luminosity function of those possible member early-type galaxies, we selected the brightest elliptical galaxies. The photometric redshifts of selected elliptical galaxies were compared with the cluster mean redshifts in the literature, and also confirmed with SDSS spectroscopic redshifts, when available. For Abell 3330 and Abell 389, the member early-type galaxies were given from the catalogs in Fasano et~al. (2000), and we selected the brightest elliptical galaxy in their Gunn $r$-band photometry.

In order to compare the optical photometry of Abell clusters together with the local samples in the Third Reference Catalog of Bright Galaxies (RC3, de Vaucouleurs et~al. 1991), we have adopted $modelMag$ from SDSS photometry and SExtractor MAG\_AUTO magnitudes in Fasano et al. (2000) catalog as the total magnitudes. The optical apertures are large enough and comparable to those in the UV photometry. And then we have transformed the optical photometry with different filter sets (Bessell $B$ \& Gunn $r$ in Fasano et~al. 2000; SDSS $g$ \& $r$) into Johnson $V$-band, by using elliptical galaxy model spectral grids (age = $5\sim12$~Gyrs) from Yi et~al. (1999). In practice, the optical spectra of model elliptical galaxies are redshifted to each target distance, and then we derived the linear forms of color--color relation by convolving the model spectra with the filter throughputs. For example, at $z=0.1$ :

\begin{displaymath}
\begin{array}{c}
(V - r) = -0.212\pm0.014 + 0.305\pm0.008\times(B - r),\\
and\\
(V - r) = -0.013\pm0.007 + 0.347\pm0.007\times(g - r).
\end{array}
\end{displaymath}

\section{RESULTS}

\begin{deluxetable*}{ccccccccl}
\tablecaption{Brightest cluster elliptical galaxies at $z < 0.2$}
\tablewidth{0pt}
\tablehead{
\colhead{Cluster} &
\colhead{SDSS} &
\colhead{RA} &
\colhead{DEC} &
\colhead{$z\tablenotemark{a}$} &
\colhead{FUV} &
\colhead{$D_{FUV}\tablenotemark{b}$} &
\colhead{NUV$\tablenotemark{c}$} &
\colhead{V} \\
\colhead{name} &
\colhead{ObjId} &
\colhead{J2000} &
\colhead{J2000} &
\colhead{} &
\colhead{} &
\colhead{} &
\colhead{} &
\colhead{} \\
\colhead{} &
\colhead{} &
\colhead{(h:m:s)} &
\colhead{(d:m:s)} &
\colhead{} &
\colhead{(mag)} &
\colhead{(arcsec)} &
\colhead{(mag)} &
\colhead{(mag)}
}
\startdata
A2399 & 587726878880694308$\tablenotemark{e}$ & 21:57:29.4 & $-$07:47:44.6 & 0.058 & 21.38$\pm$0.15 & 23.8 & 20.41$\pm$0.06 & 14.73$\pm$0.02\\
A2670 & 587727225689538702$\tablenotemark{e}$ & 23:54:13.7 & $-$10:25:08.5 & 0.078 & 20.61$\pm$0.09 & 41.6 & 19.88$\pm$0.05 & 14.45$\pm$0.02\\
A2249 & 587729782810345880$\tablenotemark{e}$ & 17:09:43.8 & $+$34:24:25.5 & 0.079 & 22.31$\pm$0.33 & 16.9 & 21.60$\pm$0.14 & 15.46$\pm$0.02\\
A2448 & 587730817899102283$\tablenotemark{e}$ & 22:31:43.2 & $-$08:24:31.7 & 0.083 & 21.35$\pm$0.20 & 23.1 & 20.45$\pm$0.05 & 14.34$\pm$0.02\\
A3330$\tablenotemark{d}$ & \nodata & 05:14:39.5 & $-$49:03:29.0 & 0.092 & 20.74$\pm$0.10 & 32.9 & 19.81$\pm$0.03 & 14.70$\pm$0.03\\
A2048 & 587729160055030043 & 15:15:14.1 & $+$04:23:10.4 & 0.097 & 22.70$\pm$0.31 & 24.0 & 21.40$\pm$0.09 & 15.44$\pm$0.02\\
A0389$\tablenotemark{d}$ & \nodata & 02:51:24.9 & $-$24:56:38.4 & 0.113 & 21.69$\pm$0.11 & 48.7 & 20.71$\pm$0.05 & 15.25$\pm$0.05\\
A0733 & 587725471208767612$\tablenotemark{e}$ & 09:01:30.1 & $+$55:39:16.7 & 0.115 & 21.89$\pm$0.19 & 18.8 & 21.42$\pm$0.12 & 15.00$\pm$0.02\\
A1406 & 587725552285122744$\tablenotemark{e}$ & 11:53:05.3 & $+$67:53:51.6 & 0.117 & 21.93$\pm$0.21 & 21.6 & 21.31$\pm$0.10 & 15.49$\pm$0.02\\
A0951 & 588017977808191595$\tablenotemark{e}$ & 10:13:50.8 & $+$34:42:51.1 & 0.143 & 22.71$\pm$0.14 & 25.7 & 22.14$\pm$0.07 & 16.34$\pm$0.02\\
A2235 & 587725993039888704 & 16:54:43.3 & $+$40:02:46.4 & 0.151 & 22.28$\pm$0.16 & 40.4 & 21.32$\pm$0.05 & 15.98$\pm$0.02\\
A1979 & 587736976342450407 & 14:50:58.7 & $+$31:17:42.0 & 0.169 & 23.07$\pm$0.13 & 17.7 & 22.65$\pm$0.07 & 16.49$\pm$0.02
\enddata
\tablenotetext{a}{SDSS spectroscopic redshift, if available. Otherwise, cluster mean redshift}
\tablenotetext{b}{$D = 2 {\times} kron\_radius {\times}{\sqrt{a\_image{\times}b\_image}} {\times}$ 1.5 arcsec/pixel}
\tablenotetext{c}{NUV magnitude measured within FUV aperture}
\tablenotetext{d}{Optical data from Fasano et~al. (2000)}
\tablenotetext{e}{Spectroscopic targets in SDSS DR4}
\label{tab:tab2}
\end{deluxetable*}

We have analyzed the brightest elliptical galaxies in 12 Abell clusters, and their photometric results are presented in Table~2. From the GALEX FUV/NUV deep and medium imaging survey data, we have measured the FUV total magnitudes and NUV flux within predefined FUV apertures. Optical V total magnitudes have also been estimated from the SDSS photometry or from the catalog data in Fasano et~al. (2000), after the filter transformation based on model elliptical galaxy spectral grids of Yi et~al. (1999), as described above. All the apparent magnitudes presented in the table are corrected for foreground extinction.

\begin{figure*}
\begin{center}
\epsscale{1.0}
{\plotone{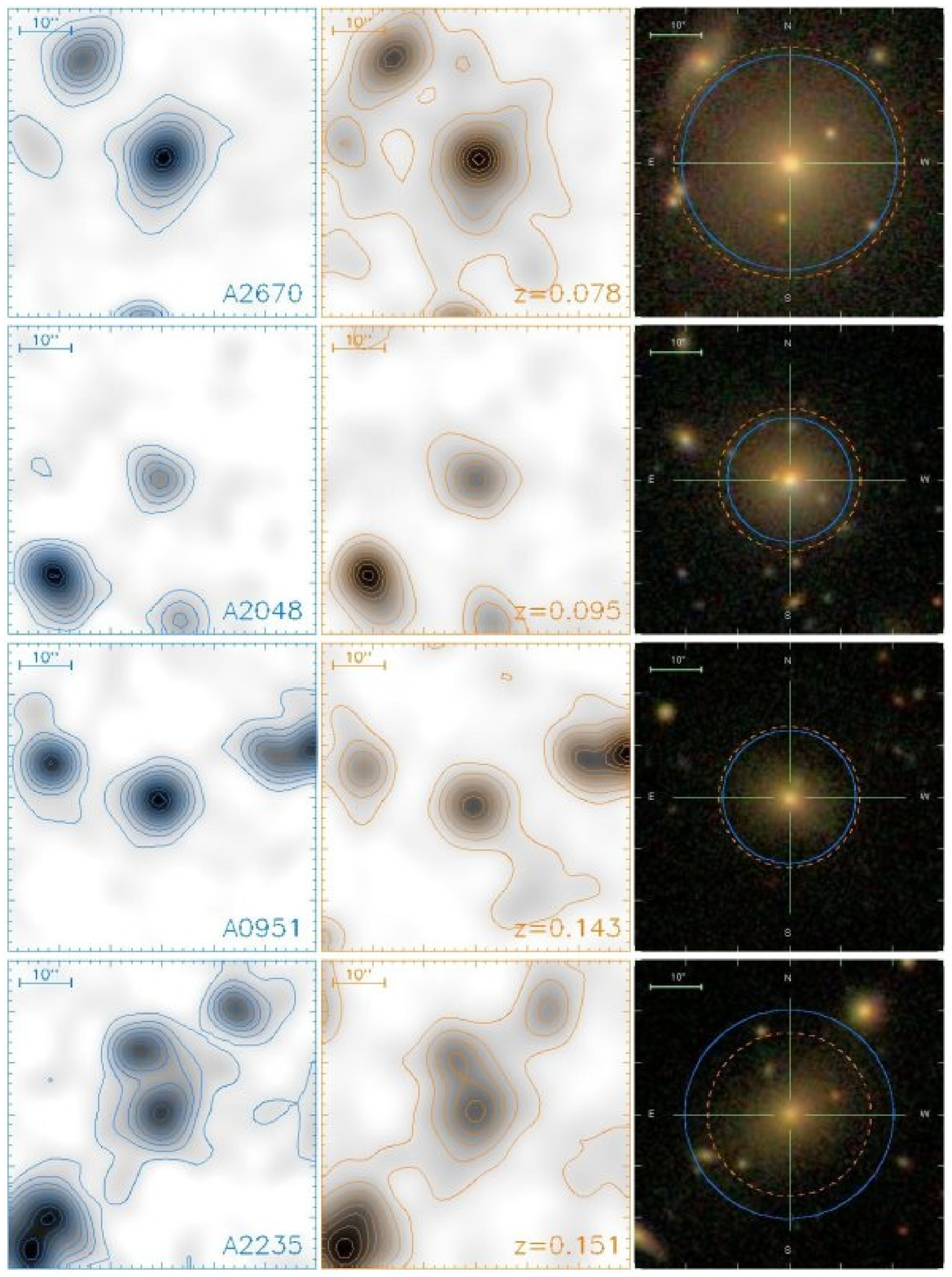}}
\end{center}
\caption{$1\arcmin\times1\arcmin$ portion of GALEX FUV/NUV ($left$/$center$) images observed in DIS mode, and SDSS optical pseudo-color images ($right$) of the brightest cluster elliptical galaxies at moderate redshifts. UV images are background-subtracted with GALEX pipeline sky maps, and FUV images are 2-pixel-smoothed for better presentation. Image contrast and the flux levels of contour maps in UV images are not consistent between the frames. Circle of solid and dashed lines in the right panel represents the FUV equivalent radius ($kron\_radius {\times}{\sqrt{a{\times}b}}$) measuring SExtractor MAG\_AUTO, and the radius containing 90\% of Petrosian flux in $r$-band ($petroR90\_r$), respectively. Note the smoothly extended UV profiles of the target ellipticals, and also UV-bright background objects which can hardly be found in optical images.\label{fig1}}
\end{figure*}

\begin{figure*}
\begin{center}
\epsscale{1.0}
{\plotone{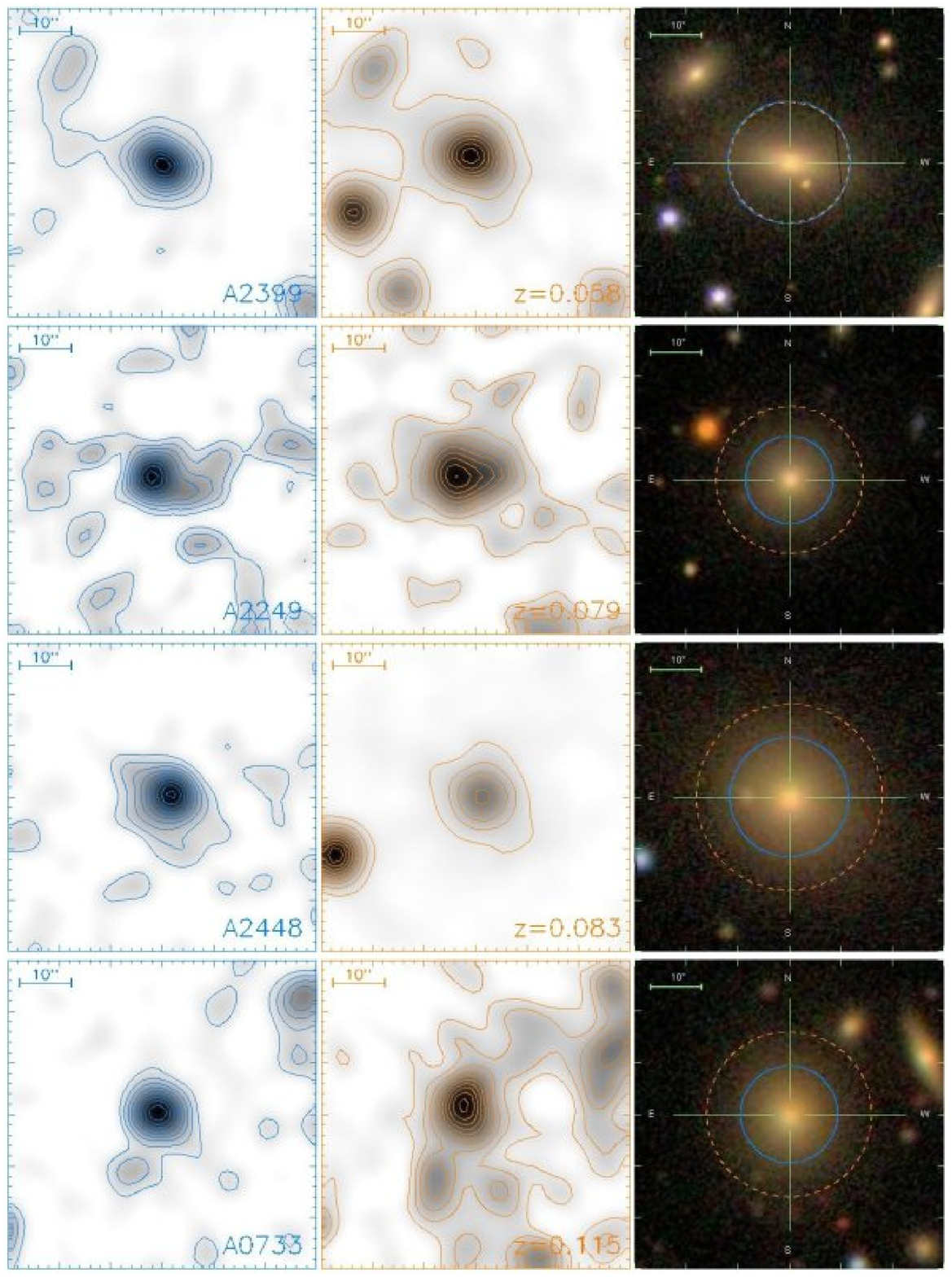}}
\end{center}
\caption{Same as Fig.~1, but for the galaxies observed in GALEX MIS mode ($\sim1.5$~ksec ; $\sim3.0$~ksec for A2399). Due to the low signal-to-noise ratio (especially in FUV) from short exposures, the extraction areas in UV images are smaller than those in optical band image, and therefore small amount of aperture effect depending on UV exposure time arises in measuring UV $-$ optical colors (see text). \label{fig2}}
\end{figure*}

\subsection{UV morphology and blending}

The FUV and NUV images of the brightest cluster elliptical galaxies are compared with the SDSS optical images in Figures 1 and 2, for those observed in GALEX DIS and MIS modes, respectively. In the rightmost panels, the MAG\_AUTO source extraction areas in the FUV images (circles of solid lines) are overlayed onto the optical images and compared with the area containing 90\% of the Petrosian flux (Petrosian 1976; Blanton et~al. 2001; Yasuda et~al. 2001) in the SDSS $r$-band images (circles of dashed lines). The UV light of the sample galaxies is centrally concentrated but also smoothly extended to the faint outermost parts, tracing the optical lights from the underlying stellar population. No indications of strong nuclear activity or massive star formation are found in the GALEX UV images or in the SDSS optical images and spectra. 

It can be seen that the outermost parts of low surface brightness in the target elliptical galaxies are not fully detected in their UV images from MIS mode (Fig.~2), mainly due to the low signal-to-noise ratio from short exposure time. This raises a varying but minor level of aperture effect in measuring UV -- optical colors, depending on the UV exposure time. We tested our DIS sample galaxies by comparing their single-orbit (only $t_{exp} >$ 1500s selected) images to the coadded, and found that the MAG\_AUTO source extraction radii in single-orbit images are $76(\pm14)\%$ (FUV) and $88(\pm25)\%$ (NUV) of those measured in DIS images. This results in $\Delta$$m(MIS - DIS)$ = $0.18(\pm0.28)$ and $0.10(\pm0.35)$ mag in FUV and NUV, respectively. Due to the limited sample size and the large errors from low countrate statistics, however, we do not apply these corrections to our MIS galaxies presented here yet. Further investigations would be required in order to describe the statistics more in detail for the photometry of such remote UV-faint galaxies.

\begin{figure*}
\begin{center}
\epsscale{0.75}
{\plottwo{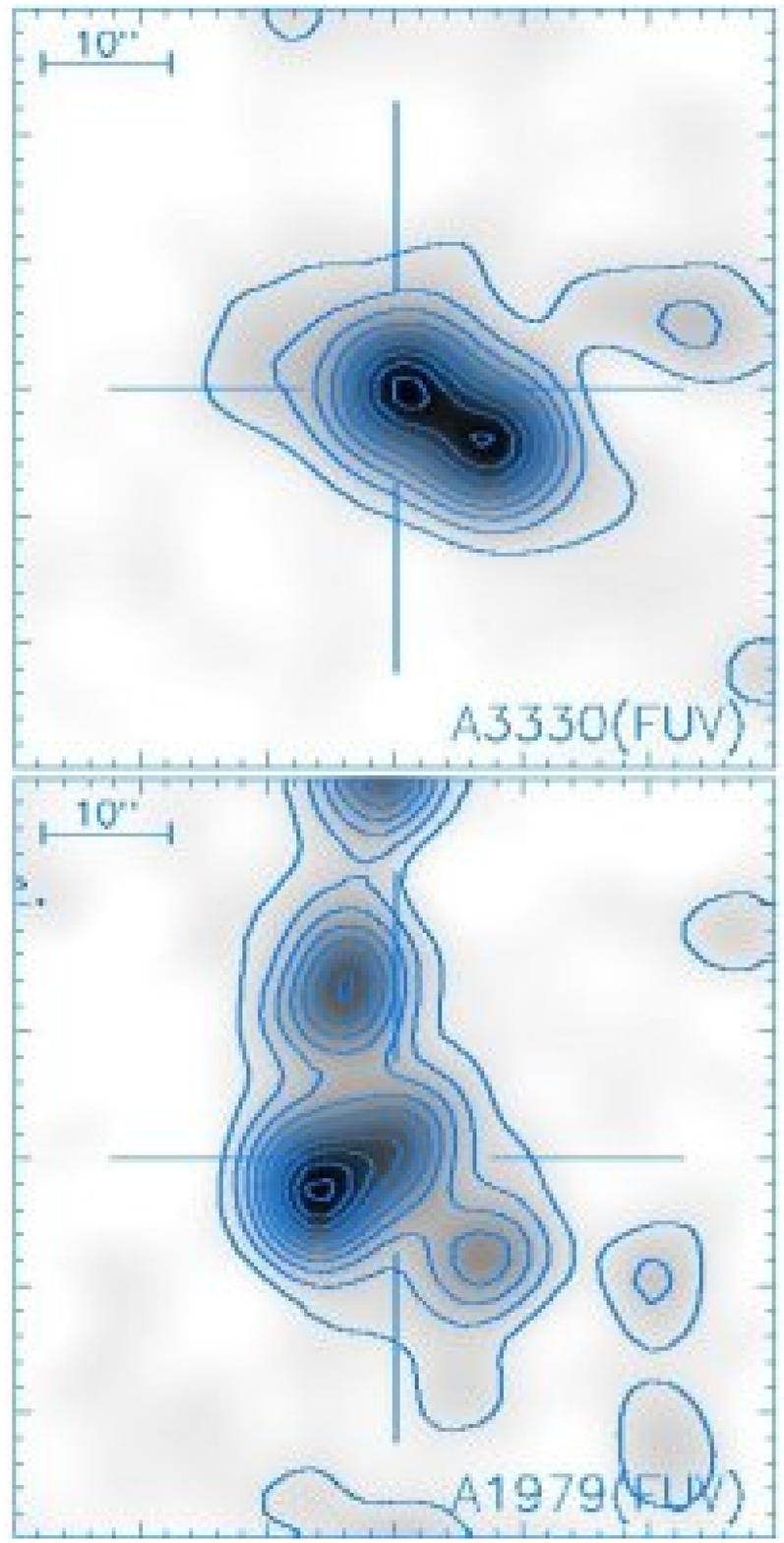}{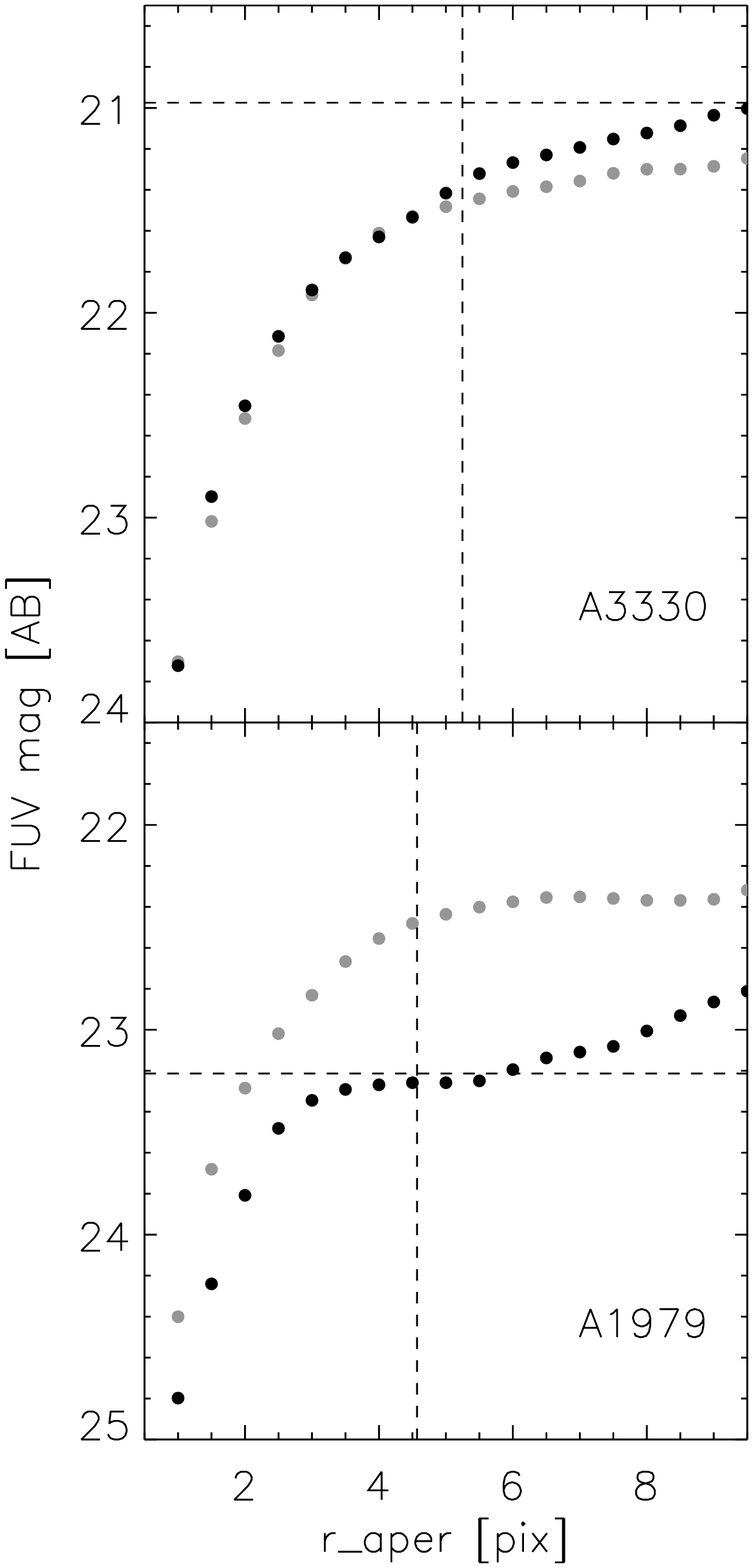}}
\end{center}
\caption{Blending in GALEX UV images. $Left$ : GALEX FUV image of the brightest elliptical galaxy in A3330 (top) and A1979 (bottom). The images are centered on the target elliptical galaxies and isophotal contours show their FUV lights are blended (or even overwhelmed) by the neighboring object. $Right$ : FUV magnitudes at different circular aperture radii. Filled circles denote the FUV growth curves of target elliptical galaxies (black) and neighbors (grey) from manual aperture photometry. Vertical dashed line indicates the separation ($\sim5$ pix) between the peak pixels of the target elliptical galaxy and the neighboring object. Horizontal dashed line indicates the total magnitudes of the target elliptical galaxy measured by SExtractor MAG\_AUTO after parameter adjustment for deblending.\label{fig3}}
\end{figure*}

For the target elliptical galaxies in A3330 and A1979, we found their UV lights blended with those from neighboring sources -- probably, star-forming dwarf companion galaxies. Figure~3 shows the FUV images (left) and growth curves (right) of the target elliptical galaxies and the neighboring objects. Even though the merged sources are deblended and detected separately after the SExtractor parameter adjustment, the FUV total magnitudes of the target elliptical galaxies measured by MAG\_AUTO (horizontal dashed lines in right panels in Fig.~3) would be more uncertain than the measured photometric errors. Due to the limited angular resolution (FWHM $\sim6\arcsec$) of GALEX UV images, the contamination from blue dwarf companion galaxies seems to be a serious issue for the study of distant massive early-type galaxies in dense environments. Multi-band optical deep imaging and spectroscopy would be necessary for the identifications of such UV contaminators.

\subsection{Optical spectra}

Although the objects studied in this paper are elliptical galaxies with typically little star forming signatures, it is important to ascertain the level of possible contamination of the UV continuum from AGN. Out of 12 brightest cluster elliptical galaxies at moderate redshifts presented here (see Table~2), 7 galaxies (A2399, A2670, A2249, A2448, A0733, A1406, and A0951) have optical spectra in SDSS DR4. While the spectra of these galaxies are inconsistent with Type I AGN, Type II (i.e., partly obscured) AGN can plausibly contribute to the UV flux. Type II AGN can be detected and distinguished from normal star-forming galaxies using the flux ratios of optical emission lines (Baldwin, Phillips, \& Terlevich 1981, BPT hereafter). Using the emission line ratios [OIII]/H$\alpha$ and [NII]/H$\beta$, Kauffmann et~al. (2003) have used a BPT-type analysis to classify a large sample of SDSS galaxies into star-forming objects and Type II AGN (Seyferts, LINERs and transition objects). We have tested for potential AGNs in our sample using the criteria derived by Kauffmann et~al. (2003).

We have utilized a code originally developed to process the spectra obtained by the SAURON project (Sarzi et~al. 2006) to measure the emission line ratios in our sample of elliptical galaxies. The code has been extensively adapted to deal with SDSS spectra and simultaneously fits the stellar continuum and emission lines. The emission lines are modelled as additional Gaussian templates. The code searches iteratively for the best velocities and velocity dispersions and solves linearly at each step for the amplitudes and the optimal combination of stellar templates in Bruzual \& Charlot (2003), which are convolved with the best line-of-sight velocity dispersion. The fitting requires that Balmer lines and forbidden lines have identical kinematics.

Based on the optical emission line analysis above, none of our 7 brightest cluster elliptical galaxies show evidence for recent star formation or AGN contamination. Their SDSS optical spectra contain no strong emission lines with $S/N > 3$: the usual criterion for active galaxies (Kauffmann et~al. 2003). We also have modelled the internal extinction of the galaxy interstellar medium, and found that the dust extinction is negligible for our target elliptical galaxies with the SDSS spectra.

\subsection{Elliptical galaxies in nearby clusters}

\begin{figure}
\begin{center}
\epsscale{1.2}
\plotone{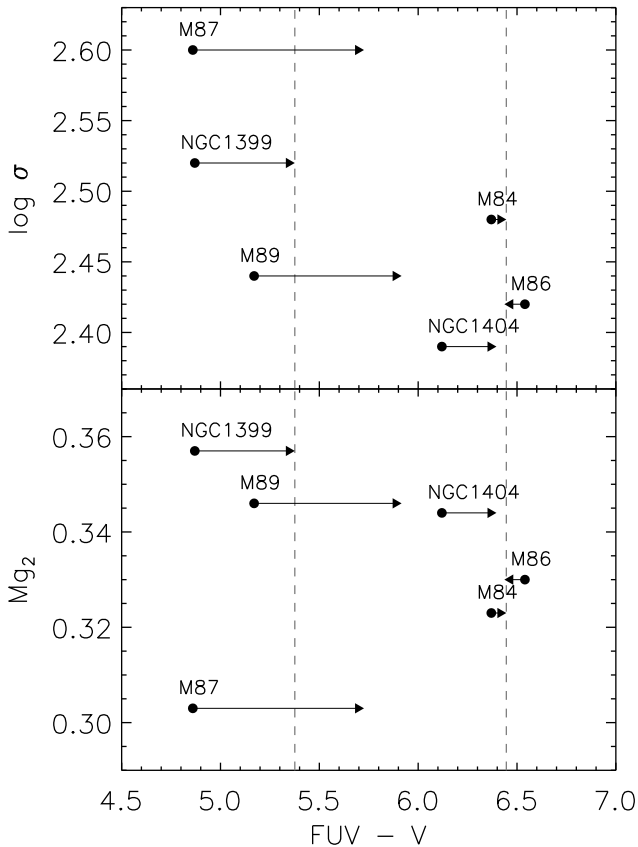}
\end{center}
\caption{Central velocity dispersion ($upper$) and Mg$_{2}$ index ($lower$) vs. $FUV - V$ color of the nearby giant elliptical galaxies in the Fornax and Virgo clusters. As the result of radial color gradient, $FUV - V$ color gets redder (except M86) when it is derived from asymptotic (total) magnitudes ($denoted~by~arrows$, data from Gil de Paz et~al. 2007), compared to the color measured within small IUE ($20\arcsec\times10\arcsec$) aperture ($filled~circles$, data from Burstein et~al. 1988). Vertical dashed lines indicate the $FUV - V$ color spread among the brightest (M$_r\lesssim-22)$ elliptical galaxies, derived from GALEX FUV and RC3 V asymptotic magnitudes.\label{fig4}}
\end{figure}

In order to compare the remote targets with the local sample, we have selected the brightest (M$_r\lesssim-22)$ elliptical galaxies in the Fornax (NGC~1399 and NGC~1404) and Virgo (M84, M86, M87, and M89) clusters. Figure~4 shows their central velocity dispersions and Mg$_{2}$ absorption line indices vs. $FUV - V$ colors. Due to the strong radial color gradient (Ohl et~al. 1998), the $FUV - V$ colors of local elliptical galaxies derived from the asymptotic (total) FUV (Gil de Paz et~al. 2007) and V (RC3) magnitudes are significantly different from those measured within small IUE ($20\arcsec\times10\arcsec$) aperture (Burstein et~al. 1988). $FUV - V$ color becomes redder (except M86) at larger aperture radii, as indicated by arrows in Figure~4, because the FUV light is more centrally concentrated in general. Apparently, the amount of color change at different aperture radii seems to be related to the strength of UV flux, in the sense that the aperture effect is larger for the galaxies with bluer nuclear $FUV - V$ colors. 

The aperture effect on $FUV - V$ colors should depend on the structural variations in the distribution of the hot component and the underlying stellar population. In comparison between the GALEX FUV surface photometry and the optical surface brightness profiles in Peletier et~al. (1990), Marcum et~al. (2001), and Lauer et~al. (2005), we found the varieties both in the FUV and the optical light distributions among the brightest elliptical galaxies of similar total luminosity in nearby clusters. M86, of which $FUV - V$ color interestingly gets bluer toward large radius from the center (see Fig.~4), has more flattened $B-$ and $R-$band surface brightness profiles (Peletier et~al. 1995) and also the distinctively lower ($\approx$ 5.7) FUV concentration parameter C31 than the other sample galaxies ($\approx$ 9) (see Table~3 in Gil de Paz et~al. 2007). Excluding M87, which has the bright nonthermal jet, NGC~1399 shows the bluest $FUV - V$ color in the sample and has one of the steepest optical radial profiles (Marcum et~al. 2001; Lauer et~al. 2005). Previous studies have found the intrinsic galaxy-to-galaxy scatter in nuclear $FUV - V$ colors (Burstein et~al. 1988) or in UV--optical color gradients (Ohl et~al. 1998) among the most massive ($\log\sigma \sim 2.5$) quiescent ellipticals in nearby clusters. The scatter becomes somewhat smaller ($\sim1$~mag, vertical dashed lines in Fig.~4) but still present when the $FUV - V$ colors are measured at much larger aperture radii of GALEX photometry.

\subsection{Look-back time evolution of FUV flux}

\begin{figure*}
\begin{center}
\epsscale{0.8}
\plotone{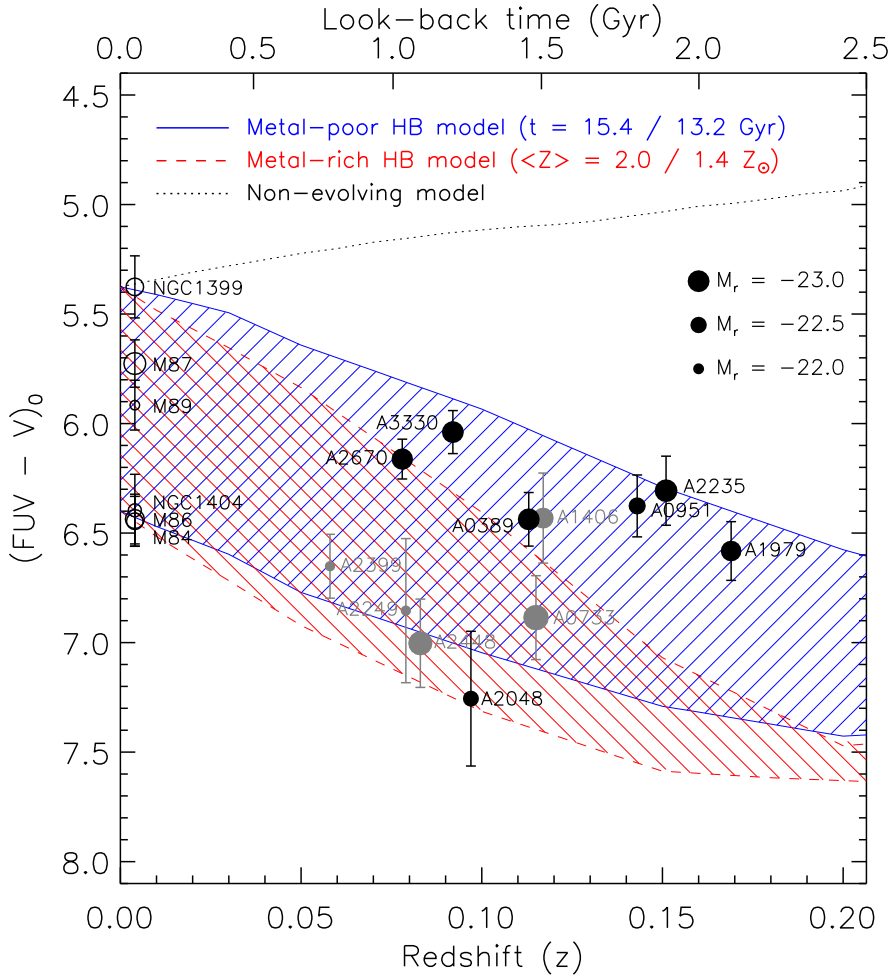}
\end{center}
\caption{Look-back time evolution of the apparent (not $K-$corrected) $FUV - V$ color for the brightest cluster elliptical galaxies at $z <$ 0.2. FUV flux fades rapidly with redshift. The colors are derived from total magnitudes to minimize aperture effect. Model lines are calibrated to the color range ($FUV - V = 5.4 - 6.4$) of the giant elliptical galaxies in nearby clusters ($open~circles$), and passively evolved and redshifted with look-back time so that they can be directly compared with the observed data of the brightest cluster elliptical galaxies ($filled~circles$) in GALEX DIS (black) and MIS (grey) mode. The size of circle symbols represents the absolute total luminosity in $r$-band. The solid and dashed lines are from the passively evolving UV-to-optical spectra of the ``metal-poor'' and ``metal-rich'' HB models, respectively. The regions filled with oblique lines denote the predicted color range from these two extreme models. The dotted line indicates the apparent color expected when NGC~1399 model spectrum is redshifted without the effect of stellar evolution. Numbers in the parentheses denote the model parameters required in each evolutionary model to reproduce the blue / red envelopes at $z=0$. See text for details.\label{fig5}}
\end{figure*}

Figure~5 shows the look-back time evolution of the apparent (not $K-$corrected) $FUV - V$ colors of the brightest cluster elliptical galaxies at $0<z<0.2$, along with the model predictions from Yi et~al. (1999). The $FUV - V$ colors are derived from the total magnitudes in both bands, in order to minimize the aperture effect that arises in the photometry for the objects at different distances, and corrected for the foreground extinction. It is clear from Figure~5, that the FUV flux from the brightest cluster elliptical galaxies fades away as redshift increases. At error weighted mean values, apparent $FUV - V$ color gets redder by 0.34 mag from local to $z\approx0.1$. For the rest-frame colors, this roughly corresponds to the spectral evolution rate of $\Delta (FUV - V) / \Delta t$ = 0.54 mag/Gyr (see $\S3.5$ for the model-based $K-$correction). The observed pace of UV fading is in good agreement with the prediction from the evolutionary population synthesis models where the mean temperature of HB stars declines rapidly with increasing look-back time. 

As seen in the local cluster elliptical samples (Fig.~4), there is also a large spread ($\sim 1$ mag) in the observed $FUV - V$ colors among the brightest cluster elliptical galaxies at $z\sim0.1$. Interestingly, the MIS sample (gray circles) appears to be redder in $FUV - V$ color than the DIS galaxies (black circles) on average. However, the observed difference between the DIS and MIS samples seems to be too large to be explained by the systematically lower FUV photon counts in MIS data which is estimated $\lesssim 0.2$ mag in $\S 3.1$. Further investigations would be required for the better cross-calibration in the GALEX UV photometry with different survey depths.

We set the ``blue'' ($\approx$ 5.4) and ``red'' ($\approx$ 6.4) envelopes in $FUV - V$ colors defined by the brightest elliptical galaxies in nearby clusters (Fig.~4), and selected the representative model spectra in the population synthesis models of Yi et~al. (1999) with two extreme assumptions for the controlling parameter that drives the UV flux. In the ``metal-poor'' HB model, age controls the FUV flux while the metallicity distribution function of the underlying population is fixed. At a given metallicity, the mean temperature of HB stars increases (and hence, FUV flux increases) with age due to the decreasing stellar envelope mass of helium-burning stars (Lee et~al. 1994; Park \& Lee 1997). In this scenario, the dominant FUV source is the old hot HB stars in metal-poor subpopulation ($Z\lesssim0.004$) which is responsible for 15~\% or less of the total optical light, and the local ($z=0$) galaxies near the blue envelope are $\sim2$~Gyr older than those near the red envelope. In contrast, in the ``metal-rich'' HB model, the FUV flux is rather controlled by the mean metallicity of the underlying stellar population. At a fixed age, the mean temperature of HB stars increases (and FUV flux increases) with increasing metallicity due to the high helium abundance and enhanced mass loss in super-metal-rich populations (Horch, Demarque, \& Pinsonneault 1992; Bressan et~al. 1994; Dorman et~al. 1995; Yi et~al. 1998). The basic assumptions and model parameters in this paper are similar to those in Yi et~al. (1999); The metal-poor and  metal-rich HB models adopt the simple (with a fixed mean metallicity of 2~Z$_\odot$) and the infall (with a fixed age of 12~Gyr for present epoch ellipticals) model metallicity distributions from Kodama \& Arimoto (1997), respectively. Readers are referred to Yi et~al. (1999) and Yoon (2002) for the details of population synthesis models.

The solid and dashed lines in Figure~5 show the apparent $FUV - V$ color evolution from the metal-poor and metal-rich models (both reach the $blue$ and $red$ envelopes at $z=0$), respectively. The model lines are passively evolved and redshifted with look-back time so that they can be directly compared with the observed data. The regions filled with oblique lines denote the predicted color ranges from the two different models. As explained in Yi et~al. (1999, see their Figure~8), the evolutionary pace of the FUV flux of elliptical galaxies at moderate redshifts ($0<z<0.2$) is predicted to be quite different between the two scenarios. This is mainly because the production of an appreciable number of hot HB stars is much more abrupt in metal-rich populations. In the metal-poor case, HB stars gradually become cooler as their masses increase with look-back time. Such an age--HB temperature relation is much more abrupt in the metal-rich case, mainly because of the large opacity effect.

\begin{figure*}
\begin{center}
\epsscale{1.0}
\plotone{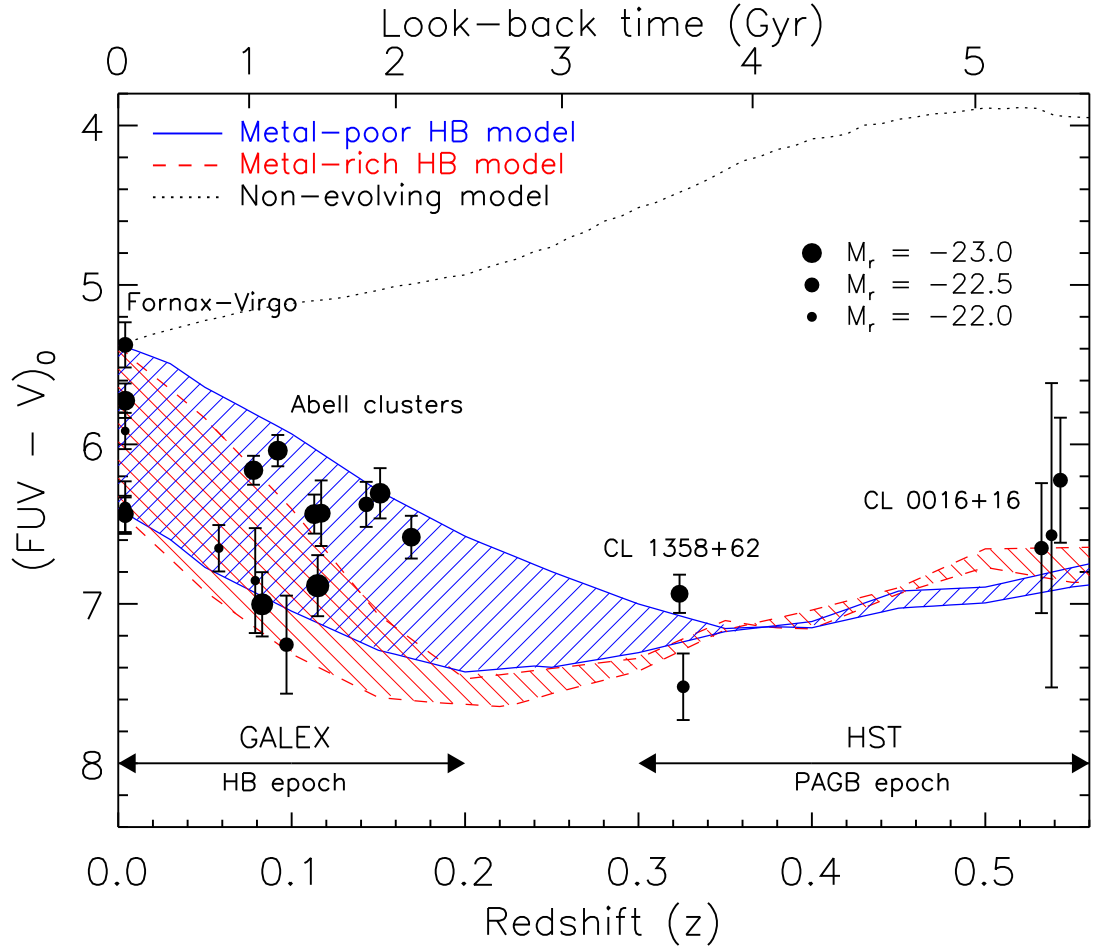}
\end{center}
\caption{Same as Fig.~5, but including the HST observations for more distant cluster elliptical galaxies (Brown et~al. 2000a, 2003). FUV flux fades away as the mean temperature of HB stars decreases for the last $\sim2$~Gyrs of look-back time. Note that the model lines converge at $z\approx0.3$ as the HB evolution effect becomes negligible. PAGB stars dominate the total FUV flux for $z>0.3$ and the adopted mass (0.565 $M_\odot$) of PAGB stars in Yi et~al. (1999) models agrees well with the observations.\label{fig6}}
\end{figure*}

In Figure~6, we compared the apparent $FUV - V$ colors of the bright giant elliptical galaxies (M$_r\lesssim-22)$ in clusters observed by the GALEX ($0<z<0.2$; this paper) and by the \textit{Hubble Space Telescope} (HST) ($0.3<z<0.6$; Brown et~al. 2000a, 2003) with the model predictions (see Lee et~al. 2005a and their Fig.~4). Again, it is clear that the FUV flux from the elliptical galaxies fades rapidly as the mean temperature of HB stars decreases for the last 2~Gyrs of look-back time. At $z\gtrsim0.3$, the HB evolution effect becomes almost negligible and the post-asymptotic giant branch (PAGB) stars dominate the total FUV flux (see also Fig.~1 of Lee et~al. 1999). The evolutionary population synthesis models of Yi et~al. (1999), based on the two extreme assumptions above, encompass the GALEX data in the lower redshift range (``HB epoch'') and agree with the HST data in the higher redshift range (``PAGB epoch'') as well (see Yi et~al. 1998 for details on the two epochs). Models which cannot account for the systematic variation of the hot stellar component (e.g., non-evolving model in Fig.~5 and 6) should be ruled out by the present observations. Readers should note that the $FUV - V$ colors presented in Figures~5 and 6 are the extinction-corrected apparent values. The non-evolving model is a model spectrum for NGC~1399 (rest-frame $FUV - V$ = 5.4), and its apparent color gets bluer with redshift because the rapidly declining optical spectrum shortward of 5000~\AA{} of an elliptical galaxy comes into Johnson $V$-band. As it is hard to presume that the optical spectra of elliptical galaxies change abruptly during the last a few Gyrs of look-back time, the observed departure from the non-evolving model line toward redder $FUV - V$ color ($> 6.0$) at moderate redshifts can only be explained with the rapid evolution of hot helium burning stars.

The observed data appear to be consistent with the variation predicted by the population synthesis models where the mean temperature of HB stars declines rapidly with increasing look-back time. However, it is hard to claim here that any one of these two extreme model assumptions should be favored by the present observations, and/or that the observed color spread among bright elliptical galaxies (both at local and at moderate redshifts) could be explained with a sole variation in either age or metallicity. There still are uncertainties in the model predictions, such as stellar mass loss and helium abundance of hot HB stars. For example, the excessively large apparent age ($\sim3$~Gyr older than the Galactic globular clusters) required for the UV-strongest local giant ellipticals in the metal-poor HB model would be reduced significantly, if the hot HB stars in giant elliptical galaxies are also explained as the minority subpopulation of super-helium-rich stars in old stellar systems as described earlier (see Lee et~al. 2005b and references therein). More detailed model construction with this effect is in progress, and therefore suffice it to say here that the two models presented in Figure~5 and 6 would bracket more realistic models to be presented later. Nontheless, it is interesting to note that the observed pace of UV fading with increasing look-back time is consistent with the systematic variations of HB temperatures, and the observed spread among remote elliptical galaxies appears to be well confined within the full range ($blue$ and $red$ envelopes) of model $FUV - V$ colors given from the passively evolving spectra of local giant elliptical galaxies.

\subsection{UV/optical color--color relation}

\begin{figure}
\begin{center}
\epsscale{1.25}
\plotone{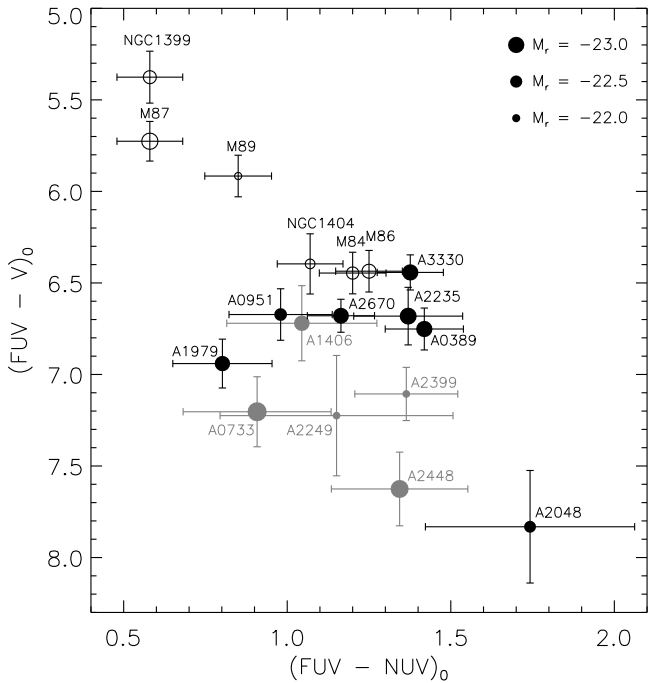}
\end{center}
\caption{Rest-frame $FUV - V$ color (derived from total magnitudes) vs. $FUV - NUV$ color (measured within FUV aperture) for the brightest cluster elliptical galaxies at $0<z<0.2$. Significant amount of evolution effect in both colors is found between the brightest elliptical samples in nearby clusters ($open~circles$) and in remote clusters ($filled~circles$) from GALEX DIS (black) and MIS (grey) images. The observed colors are corrected for the Galactic foreground extinction, and converted to the rest-frame values with a template model spectrum of UV upturn elliptical galaxy that has apparent $FUV - V$ $\simeq$ 6.5 at $z$ = 0.1. Aperture correction of $\Delta(FUV - NUV)=+0.15$ is applied to remote galaxies, for the effects of $FUV - NUV$ color gradient, in order to be compared with the local giant elliptical galaxies of comparable optical luminosities in the Fornax and Virgo cluster (asymptotic colors from Gil de Paz et~al. 2007). \label{fig7}}
\end{figure}

Figure~7 shows the rest-frame color--color diagram, $FUV - V$ vs. $FUV - NUV$, for the brightest cluster elliptical galaxies at $0<z<0.2$. In order to be compared with the local sample, the colors of remote target galaxies are $K-$corrected with a template UV upturn model spectrum from Yi et~al. (1999) that has apparent $FUV - V \simeq 6.5$ (the error weighted mean value of observed data) at $z = 0.1$. It can be seen that a tight color--color relation can be defined by the local giant elliptical sample in nearby clusters, in the sense that an elliptical galaxy which emits stronger FUV flux has more steeply increasing spectral energy with decreasing wavelength from NUV (2271\AA) to FUV (1528\AA). Such a relationship is supported by the remote elliptical galaxies, yet it is scattered as they are not at the equal cosmological distance or age. 

Even though there is an uncertainty in model-based $K-$corrections for the UV spectra of remote elliptical galaxies, it is clear that there is a significant amount of evolution with look-back time in both $FUV - V$ (a measure of UV upturn strength) and $FUV - NUV$ (a measure of UV spectral shape) colors from the local to the moderate redshifts ($0.05<z<0.17$). Taking the error weighted mean values for the local and distant samples, we derive rest-frame color evolution of $\Delta(FUV - V) = +0.74$ and $\Delta(FUV - NUV) = +0.31$, which are consistent with the model predictions.

It is worth mentioning the unusually-weak UV emission observed in the brightest elliptical galaxy in A2048 at $z\approx0.1$. Its distinctively red colors in both $FUV - V$ and $FUV - NUV$ (Fig.~7) might be caused in part by heavy internal extinction. Otherwise, its observed colors indicate that the age of the oldest stellar population therein would be much less than 10~Gyr, as inferred from UV/optical spectral grids of elliptical galaxy models in Yi et~al. (1999).

\section{DISCUSSION}

We have analyzed the UV images of the brightest (M$_r\lesssim-22)$ elliptical galaxies in 12 Abell clusters from the ongoing GALEX imaging survey for the elliptical-rich clusters at moderate redshifts. Their photometric results are compared with the local giant elliptical galaxies of comparable optical luminosity in the Fornax and Virgo clusters. The sample galaxies appear to be quiescent without signs of massive star formation or strong nuclear activity, and show smooth, extended UV profiles indicating that the FUV light is likely produced by the hot stars in the underlying old stellar population. As a measure of the UV upturn strength, the $FUV - V$ colors of remote target galaxies are compared with the local giant ellipticals of comparable optical luminosity. We have found that the FUV flux from the brightest cluster elliptical galaxies fades away with look-back time at moderate redshifts ($0<z<0.2$), and the observed pace of the UV fading is in good agreement with the prediction from the population synthesis models where the mean temperature of HB stars declines rapidly with increasing look-back time.

The GALEX data presented here and in Lee et~al. (2005a) show the rapidly fading UV spectra of elliptical galaxies over the past 2~Gyrs of look-back time, being consistent with the prediction that the FUV flux from quiescent elliptical galaxies is mainly produced by hot HB stars and their progeny. They imply a strong age dependency in the evolution of the FUV sources responsible for the UV upturn phenomenon in old stellar systems. It is unlikely that the episodic star formation or AGN outburst reproduces the systematic fading in UV flux with look-back time, although their contamination on the UV flux is evident in some cases (e.g., O'Connell et~al. 2005; Rich et~al. 2005; Yi et~al. 2005; Schawinski et~al. 2006). Various hot subdwarf formation mechanisms in binary systems may also contribute to the UV flux in early-type galaxies; yet such models appear to have difficulty reproducing the redshift evolution of the UV upturn found so far (Brown et~al. 2006; Han et~al. 2006).

The observed spread ($\sim1$ mag) in $FUV - V$ (and also in $FUV - NUV$) color among the giant elliptical galaxies in nearby clusters is also present among the brightest cluster ellipticals at around $z=0.1$. Although the origin of such a color spread among the most massive ellipticals at equal cosmological distances is still to be understood, the observed spread among remote elliptical galaxies ($0.05<z<0.17$) appears to be encompassed with the full range of model $FUV - V$ colors ($blue$ and $red$ envelopes) set by the passively-evolving spectra of local giant elliptical galaxies having $5.4 \lesssim FUV - V \lesssim 6.4$ at $z=0$.

Whether the strong age dependency of the properties of FUV sources and the observed spread in $FUV - V$ color imply the age dispersion among the local giant elliptical galaxies (e.g., Park \& Lee 1997; Greggio \& Renzini 1999) requires further investigation, due to some current technical difficulties in observational analysis (e.g., blending, aperture effect, internal reddening) and to the uncertainties in theoretical interpretation (e.g., super-helium-rich extreme HBs, $\Delta$Y/$\Delta$Z, metallicity distribution function). Nonetheless, the GALEX UV photometry has confirmed that the UV upturn is  the most rapidly evolving feature in old stellar systems and thus eventually should be used as a good age indicator (Greggio \& Renzini 1990; Park \& Lee 1997; Yi et~al. 1999). Planned observations with GALEX for more cluster targets at $0<z<0.25$ would certainly help us derive the stellar ages of elliptical galaxies and constrain their formation epoch through empirical approach. The ongoing search for the correlation between the UV strength and other physical parameters (such as velocity dispersion and H$\beta$ index) among the cluster and field early-type galaxies with a wide range of luminosity funcion at each redshift bin, and its dependency on the environment are highly anticipated as well.

\acknowledgments{
GALEX (Galaxy Evolution Explorer) is a NASA Small Explorer, launched in April 2003. We gratefully acknowledge NASA's support for construction, operation, and science analysis for the GALEX mission, developed in cooperation with the Centre National d'Etudes Spatiales of France and the Korean Ministry of Science and Technology. Yonsei University participation was supported by the Creative Research Initiative Program of MOST/KOSEF. We are grateful to Marc Sarzi for providing his spectral fitting code, adapted for use on SDSS spectra. S.K.Y. acknowledges support by grant No. R01-2006-000-10716-0 from the Basic Research Program of the Korea Science \& Engineering Foundation.}

\end{document}